\documentclass[preprint]{aastex}

\shorttitle{Gas removal in clusters}
\shortauthors{F. C. Adams}

\newcommand{\be}{\begin{equation}}
\newcommand{\ee}{\end{equation}}
\newcommand{\ep}{\varepsilon} 
\newcommand{\rhostar}{{\rho_\ast}} 
\newcommand{\rhogas}{{\rho_{\rm gas}}} 
\newcommand{\erf}{{{\rm Erf}}} 
\newcommand{\ewig}{{{\widetilde \epsilon}}} 
\newcommand{\rt}{{r_T}} 
\newcommand{\atan}{{\rm atan}}

\newcommand{\vm}{{v_{\rm m}}}
\newcommand{\xit}{{\xi_T}}

\newcommand{\muck}{{\mu_C}} 
 
\newcommand{\fun}{ {\cal F}_{\ast}}

\begin{document}

\title{Theoretical Models of Young Open Star Clusters: 
\\ Effects of a Gaseous Component and Gas Removal}

\author{Fred C. Adams} 
 
\affil{Physics Department, University of Michigan, Ann Arbor, MI 48109}

\email{fca@umich.edu} 

\begin{abstract} 

We construct a family of semi-analytic models for young open clusters,
including a gaseous component and varying assumptions about the
distribution function for the stellar component. The parameters of
these models are informed by observed open clusters and general
theoretical considerations regarding cluster formation.  We use this
framework to estimate the fraction $\fun$ of the stellar component
that remains gravitationally bound after the gaseous component
disperses. The remaining stellar fraction $\fun$ is a smooth function
of the star formation efficiency $\epsilon_\ast$, and depends on the
distribution function of the stars.  We calculate the function
$\fun(\epsilon_\ast)$ for this class of open cluster models and 
provide fitting formulae for representative cases. 
 
\end{abstract}

\keywords{open clusters and associations: general -- stars: formation} 

\section{INTRODUCTION} 

In this paper, we study the equilibrium structure and early evolution
of young open star clusters. Our long term goal is to develop a
unified treatment that includes cluster formation, removal of the
gaseous component, and the longer term evolution of the system.  A
unified picture of young cluster evolution can be useful in several
different contexts. On one hand, we can consider an open cluster as an
astrophysical entity and study its birth, evolution, and ultimate
demise. On the other hand, we can study the effects of the cluster
setting on the star formation process. In this present work, we
construct models for young clusters in which a gaseous component is
still present. We then study how the clusters react to gas removal.
Our results are applicable to robust clusters that live for relatively
long times ($\sim100$ Myr); perhaps 10\% of all star formation takes
place in such robust cluster environments (see our companion paper
Adams \& Myers 2000).

A collection of previous papers have addressed the question of whether
or not a cluster can remain gravitationally bound after the removal of
its gaseous component (e.g., Hills 1980; Mathieu 1983; Elmegreen 1983;
Lada, Margulis, \& Dearborn 1983).  Most previous work uses an
approach based on the virial theorem, however, and does not explicitly
include the distribution function for the stars.  In the case of rapid
gas removal and a cluster that begins in a state of exact virial
equilibrium, the cluster expands by a factor $f_{\rm ex}$ after the 
gas is removed. In the simplest models, this factor $f_{\rm ex}$ is 
given by 
\be 
f_{\rm ex} = {\epsilon_\ast \over 2 \epsilon_\ast - 1} \, , 
\label{eq:fsimple} 
\ee
where $\epsilon_\ast$ is the mass fraction of the stellar component in
the original cluster. In this approximation, the cluster remains bound
for $\epsilon_\ast > 1/2$; as $\epsilon_\ast \to 1/2$, 
$f_{\rm ex} \to \infty$ and the cluster becomes unbound. This approach
has been generalized to include varying assumptions about the speed of
gas removal and departures from exact virial equilibrium, e.g.,
stellar speeds that are less then the virial speeds in the cluster
potential (see, e.g., Verschueren 1990; Verschueren \& David 1989;
Elmegreen \& Clemens 1985; Lada et al. 1983).  In this present work,
we take into account the distribution function for the stars and,
separately, the density distribution of the gaseous component.  These
complications lead to a much wider range of possible behavior.
Because the cluster stars have a velocity distribution, the low speed
stars in the tail of the distribution survive as a gravitationally
bound entity even if $\epsilon_\ast < 1/2$; the high velocity stars in
the opposite tail escape even if $\epsilon_\ast >$ 1/2. For a given
distribution function and a given star formation efficiency
$\epsilon_\ast$, a fraction $\fun (\epsilon_\ast)$ of the stars will
thus remain bound after gas removal. The function $\fun
(\epsilon_\ast)$ varies smoothly with star formation efficiency rather
than exhibiting singular behavior.

This paper is organized as follows. In \S 2, we construct equilibrium
models of young open clusters including standard forms for the stellar
distribution function, a gaseous component, and anisotropy parameters.
We study the effects of gas removal in \S 3; in particular, we
calculate the fraction $\fun$ of stars remaining after gas leaves the
system.  We compare our results with observed clusters in \S 4 and
then conclude, in \S 5, with a summary and brief discussion of our
results. In the Appendix, we also present a crude cluster formation
theory, which informs the theoretical models in the main text.
Starting with the existing theory for the collapse of molecular cloud
cores, we scale up the solutions to describe the collapse flow that
forms a star cluster.  Using this formalism, we estimate the
anisotropy of the velocity distribution for forming clusters.

\section{EQUILIBRIA OF STAR CLUSTERS CONTAINING GAS} 

In this section, we construct a class of cluster models that
incorporate both stars and gas. In this standard approach, the
structure of clusters is determined by two differential equations. The
first is the Poisson equation for the gravitational potential $\Phi$, 
\be 
\nabla^2 \Phi = 4 \pi G \, \rho_{\rm tot} \, , 
\label{eq:poisson} 
\ee
where the total density $\rho_{\rm tot}$ includes both stars and gas. 
The second is the collisionless Boltzmann equation, which takes the form 
\be
{\partial f \over \partial t} + {\bf v} \cdot {df \over d{\bf x}} 
- \nabla \Phi \cdot {df \over d{\bf v}} = 0 \, , 
\label{eq:boltz} 
\ee
where $f$ = $f({\bf x}, {\bf v}, t)$ is the distribution function 
for the stars. 

\subsection{Models with isotropic velocity dispersion} 

As a starting point, we construct equilibrium models with an isotropic
velocity dispersion. Making straightforward extensions of the general
treatment outlined in Binney \& Tremaine (1987), we include a gaseous
halo in the formalism. In particular, we use lowered isothermal
cluster models by assuming a stellar distribution function $f(\ep)$ 
of the form 
\be 
f(\ep) = \rho_1 (2 \pi \sigma^2)^{-3/2} 
\bigl[ {\rm e}^{\ep/\sigma^2} - 1 \bigr] \, , 
\label{eq:distfun} 
\ee 
where the relative energy $\ep$ is defined by $\ep = \Psi - v^2/2$,
and where the relative potential $\Psi$ is related to the true
gravitational potential $\Phi$ through $\Psi = - \Phi + \Phi_0$. 
The constant $\rho_1$ sets the density scale and the parameter
$\sigma$ sets the scale for the velocity dispersion (notice that 
$\sigma \ne \langle v^2 \rangle^{1/2}$).  By assuming a distribution
function which is a function of the energy only, we automatically
obtain a solution to the collisionless Boltzmann equation
[\ref{eq:boltz}] for equilibrium configurations with $\partial f$/
$\partial t$ = 0. We have also implicitly assumed an isotropic
distribution in velocity space -- we consider the more general 
case of anisotropic models in a subsequent section. 

From the distribution function $f(\ep)$, we can determine the stellar
density $\rhostar$ in terms of the potential. Using the resulting
density in the Poisson equation, in conjunction with the additional
terms resulting from gas, we can find the potential as a function of
radius in the cluster, and then find the stellar density as a function
of radius.  For convenience, we define new variables
\be 
\psi \equiv \Psi / \sigma^2 \, , \qquad 
\xi \equiv r/r_0 , \qquad {\rm and} \qquad 
r_0 = \bigl[ 9 \sigma^2 / 4 \pi G \rho_0 \bigr]^{1/2} \, , 
\ee
where $\rho_0$ is the central density of the {\it stellar} component
of the cluster.  The variable $r_0$ is thus the King radius and 
plays the role of an effective core radius for the stellar component 
(see Binney \& Tremaine 1987).

Using the newly defined variables, we can evaluate the density of
stars $\rhostar (r)$ in terms of the reduced potential $\psi (r)$,
i.e., 
\be 
\rhostar = \int f d^3 v = \rho_1 \Bigl[ {\rm e}^\psi 
\erf ( \sqrt{\psi} ) - 2 \sqrt{ {\psi/\pi}} 
\bigl( 1 + {2 \over 3} \psi \bigr) \Bigr] \, , 
\label{eq:defrhostar} 
\ee 
where $\erf (z)$ is the error function (e.g., Abramowitz \& Stegun 
1972). The central density can thus be written 
\be 
\rho_0 = \rho_1 \Bigl[ {\rm e}^{\psi_0} 
\erf ( \sqrt{\psi_0} ) - 2 \sqrt{ {\psi_0/\pi}} 
\bigl( 1 + {2 \over 3} \psi_0 \bigr) \Bigr] \, , 
\ee 
where $\psi_0$ = $\psi(0)$ = $\Psi(0)/\sigma^2$ determines the depth
of the gravitational potential well at the cluster center.  We find 
it convenient to define a function 
\be 
g_\ast (x) \equiv {\rm e}^x \, \erf ( \sqrt{x} ) - 
2 \sqrt{ {x \over \pi}} \, (1 + 2x/3) \, , 
\label{eq:gfun} 
\ee
so that $\rhostar = \rho_1 \, g_\ast(\psi)$ = 
$\rho_0 \, g_\ast(\psi)/g_\ast(\psi_0)$. 

To specify the gas contribution, we use a density distribution much
like an isothermal sphere; in particular, we consider a gaseous halo
around the cluster with a density profile of the form 
\be 
\rhogas = {a^2 \over 2 \pi G (r^2 + r_{\rm 0g}^2) } \, , 
\label{eq:gashalo} 
\ee
where $a$ is the effective sound speed of the gas and where 
$r_{\rm 0g}$ is the core radius for the gas. For simplicity, 
we assume that the gas and the stars have the same core radius 
$r_0$; in principle, we could include an additional parameter 
$\alpha_{{\rm g}\ast} \equiv r_{\rm 0g}/r_0$ in the analysis. 

The new version of the Poisson equation (which determines the 
cluster structure) takes the form 
\be 
\xi^{-2} {d \over d\xi} \Bigl( \xi^2 {d \psi \over d\xi} \bigr) 
= - {9 \over g_\ast (\psi_0) } \Bigl[ {\rm e}^\psi \erf (\sqrt{\psi}) - 
2 \sqrt{ {\psi/\pi}} \bigl( 1 + {2 \over 3} \psi \bigr) \Bigr] 
- {\Lambda  \over 1 + \xi^2} \,  , 
\label{eq:diffeq} 
\ee
where the constant $\Lambda$ determines the contribution 
of the gas relative to that of the stars and is defined by 
\be 
\Lambda \equiv {2 a^2 \over \sigma^2 } \, . 
\ee
We thus have a two parameter family of cluster models. To 
specify each model, we need to choose the value of $\psi_0$ 
to set the depth of the gravitational potential well and 
the value of $\Lambda$ to set the gas contribution.  

As we integrate equation [\ref{eq:diffeq}] outwards, both the
potential $\psi$ and the stellar density $\rhostar$ decrease and
eventually become zero at some outer truncation radius $\rt$. [Notice
that the stellar density approaches the form $\rho_\ast (\psi)$ $\sim$
($8/15\sqrt{\pi}$) $\psi^{5/2}$ in the limit $\psi \to 0$.]  At this
outer boundary, the true gravitational potential $\Phi$ is given by
$\Phi(\rt) = - G M(\rt)/\rt$, where $M(\rt)$ is the total mass
enclosed, including both stars and gas.  The value of the true
potential at the origin is then $\Phi(0) = \Phi(\rt) - \Psi(0)$, where
$\Psi(0)$ = $\psi_0 \sigma^2$. As the input parameter $\psi_0$ grows
larger, the outer truncation radius $\rt$ grows accordingly; the mass
$M(\rt)$ also increases, and so does the overall depth of the
potential well as determined by $|\Phi(0)|$.

A typical density profile is shown in Figure 1a (for a model using
$\psi_0$ = 10 and $\Lambda$ = 1/2). For this case, the mass in stars
and the mass in gas are roughly comparable (specifically, 44\% stars
and 56\% gas). The dashed curve shows the stellar component
(eq. [\ref{eq:defrhostar}]) and the dotted curve shows the gaseous
component (eq. [\ref{eq:gashalo}]). The total density is given by the
solid curve. After the gas is removed, the remaining stellar component
contains $\sim73\%$ of the starting inventory, with its resulting
density distribution depicted by the dot-dashed curve (the formulation
of this gas removal calculation is presented \S3).  For this class of
models, the density in the cluster center is dominated by stars and
the outside is dominated by gas; the star formation efficiency thus
increases toward the inside.  This structure is roughly consistent
with that expected from a collapse model of cluster formation as
outlined in the Appendix. Even though the density is strongly peaked
toward the center, most of the mass in the cluster -- in both stars
and gas -- resides in the outer portion of the cluster.  Figure 1b
shows a collection of density profiles with varying depths of the
gravitational potential, determined by $\psi_0$ which lies in the
range 6 $\le \psi_0 \le$ 14.  Similarly, Figure 1c shows a collection
of density profiles for $\psi_0$ = 10 and varying amounts of gas,
determined by $\Lambda$ which lies in the range 0 $\le \Lambda \le$ 2.

In this present construction, the stars live in a potential well
determined by both their own self-gravity and by the gravity of the
accompanying gas. The stars have a distribution function which was
chosen to be a function of an integral of motion (the energy). Through
the Jeans Theorem, the distribution function is thus a valid solution
to the collisionless Boltzman equation which determines equilibrium
configurations of stellar dynamics (see Binney \& Tremaine 1987).
Although physically motivated, the gaseous halo distribution
[\ref{eq:gashalo}] was an arbitrary choice and will not in general be
in perfect hydrostatic equilibrium with the potential obtained in
solving equation [\ref{eq:diffeq}]. To address this potential
discrepancy, we assume that support for the gaseous halo is given by
additional pressure terms produced by external agents such as magnetic
fields. This approach is adopted in the present version of the
calculation. In principle, we could also solve self-consistently for
the density distribution of the gas through an iterative procedure.

\subsection{Models with anisotropic velocity dispersion} 

In this section, we consider the more realistic case of anisotropic
models.  For an anisotropic velocity dispersion tensor, the
distribution function $f$ is a function of both the energy $\ep$ and
the angular momentum $\bf L$. To obtain a natural extension of the
formulation presented in the previous section, we use a distribution
function of the form 
\be 
f(\ep, L) = \rho_1 (2 \pi \sigma^2)^{-3/2} \, 
g (L^2 / r_A^2 \sigma^2) \, 
\bigl( {\rm e}^{\ep / \sigma^2} - 1 \bigr) \, , 
\label{eq:dfani} 
\ee
where the function $g$ and the anisotropy radius $r_A$ determine the
degree of departure from isotropy. In the limit $r_A \to \infty$, $g
\to 1$ and we recover the isotropic models of the previous section.
In our cluster formation scenario (see the Appendix), the anisotropy
radius $r_A$ is determined by the centrifugal radius $R_C$ of the
collapse flow so that $r_A \approx R_C$ (where $R_C$ is evaluated at
the end of the cluster formation epoch).

With a distribution function of the form [\ref{eq:dfani}], 
the stellar density can be written 
\be
\rhostar = \rho_1 {\sqrt{2 \over \pi}} 
\int_0^{\pi/2} \, \sin\eta \, d\eta \, 
\int_0^{\sqrt{2\psi}} \, v^2 \, dv \, 
g(r^2 v^2 \sin^2 \eta / r_A^2) \,  
\bigl( {\rm e}^{\psi - v^2/2} - 1 \bigr) \, , 
\ee 
where we have scaled out the $\sigma$ dependence. Motivated by our
model of cluster formation from a collapse flow, we use an anisotropy
function $g(L^2)$ that decreases as one power of $r$ (and hence $L$)
for large radii. We thus choose the particular form 
\be
g(L^2) = {1 \over (1 + L^2/r_A^2)^{1/2}} = 
{1 \over \bigr( 1 + r^2 v^2 \sin^2\eta /r_A^2 \bigl)^{1/2}} \, = 
{1 \over \bigr( 1 + \alpha^2 v^2 \sin^2\eta \bigl)^{1/2}} \, , 
\ee
where we have defined an anisotropy parameter $\alpha \equiv r / r_A$. 
With this choice for $g(L^2)$, the angular integral becomes 
\be
\int_0^{\pi/2} \, {\sin\eta \, d\eta \over 
\bigr[ 1 + \alpha^2 v^2 \sin^2\eta \bigl]^{1/2}} \, =
{1 \over \alpha v} \, \atan (\alpha v) \, .
\ee
The stellar density takes the form 
\be
\rhostar = \rho_1 \sqrt{2 \over \pi} {r_A \over r} 
\, \int_0^{\sqrt{2 \psi}} \, v \, dv \, 
\atan(\alpha v) \bigl( {\rm e}^{\psi - v^2/2} - 1 \bigr) \, 
\equiv \rho_1 \sqrt{2 \over \pi} {r_A \over r} \, I_\rho \, , 
\label{eq:rhostardef} 
\ee
where we have defined the integral $I_\rho$ in the final equality. 
The integral $I_\rho$ can be partially evaluated to obtain 
\be
I_\rho = { \sqrt{2 \psi} \over 2 \alpha} - \atan(\alpha \sqrt{2 \psi}) 
\bigl[ 1 + \psi + {1 \over 2 \alpha^2} \bigr] + \alpha 
{\rm e}^{\psi} \, K_\rho \, , 
\label{eq:intdef} 
\ee
where the remaining integral $K_\rho$ is given by 
\be
K_\rho (\psi, \alpha) \equiv \int_0^{\sqrt{2 \psi}} 
{{\rm e}^{-v^2/2} \over 1 + \alpha^2 v^2} \, dv \, . 
\label{eq:kintdef} 
\ee
Although the integral $K_\rho$ cannot be evaluated in terms of
elementary functions, we can obtain a good approximation using
asymptotic analysis (Bleistein \& Handelsman 1986).  In the limit 
of small $\alpha$, we invoke Laplace's method and keep only the 
leading order term to find the following analytic expression 
\be
K_\rho \approx \sqrt{\pi \over 2} (1 + 2 \alpha^2)^{-1/2} 
\, \erf \bigl[ \sqrt{\psi (1 + 2 \alpha^2)} \bigr] \, . 
\label{eq:kintsolve} 
\ee 
This approximation becomes exact in the limit of small $\alpha$ (small
departures from isotropy). In the opposite limit of large $\alpha$,
\be 
K_\rho \approx \atan(\alpha \sqrt{2\psi})/\alpha \, . 
\label{eq:kintsolve2}
\ee
The stellar density contribution is thus given by the combination of
equations [\ref{eq:rhostardef} -- \ref{eq:kintsolve2}]. Since we have
made an approximation in evaluating the density (equations
[\ref{eq:kintsolve}, \ref{eq:kintsolve2}]), the distribution function
corresponding to this density profile is not exactly of the form given
by equation [\ref{eq:dfani}], although it is close and it has the
correct behavior in the limiting regimes.

The form of the Poisson equation, which determines the potential, 
now takes the form 
\be
\xi^{-2} {d \over d\xi} \Bigl( \xi^2 {d \psi \over d\xi} \bigr) 
= - {9 h_\ast (\psi) \over h_\ast (\psi_0) } - 
{\Lambda  \over 1 + \xi^2} \,  , 
\label{eq:poisson2} 
\ee
where the function $h_\ast (x)$ is defined by 
\be
h_\ast (x) \equiv {\rm e}^x K_\rho (x, \alpha) + 
{ \sqrt{2x} \over 2 \alpha^2} - {1 \over \alpha} 
\atan(\alpha \sqrt{2x}) 
\Bigl[ 1 + x + {1 \over 2 \alpha^2} \Bigr] \, . 
\label{eq:hstardef} 
\ee
The parameter $\alpha = r/r_A$ = $\xi r_0 / r_A$, so we define an
overall anisotropy parameter $\beta \equiv r_0/r_A$. Our scenario for
cluster formation indicates that both the anisotropy radius $r_A$ and
the cluster core radius $r_0$ are approximately given by the
centrifugal radius $R_C$ from the collapse; we thus expect $r_A
\approx R_C \approx r_0$ and hence $\beta \approx 1$.  Notice that in
the limit $r \to 0$, $\alpha \to 0$, and the function $h_\ast$ takes
on a limiting form $h_\ast (x; \alpha \to 0)$ = $\sqrt{\pi/2}\,$ 
${\rm e}^x \, \erf \bigl[ \sqrt{x} \bigr] - \sqrt{2x} (1 + x)$. 

Figure 2 shows the density profiles for a representative cluster with
an anisotropic velocity distribution, where the parameters are taken
to be $\psi_0$ = 10, $\Lambda$ = 1/2, and $\beta$ = 1. The dashed
curve shows the stellar component, the dotted curve shows the gaseous
component, and the solid curve shows the total density; the dot-dashed
curve shows the reduced stellar density profile after gas removal (see
the following section). For this class of models, the velocity
anisotropy leads to a more extended stellar component compared to the
isotropic models of the previous section. For the same values of
$\psi_0$ and $\Lambda$, the stellar component extends farther into the
gaseous halo and the mass fraction of stars is correspondingly
smaller.  For this typical model using $\psi_0$ = 10 and $\Lambda$ = 1/2, 
for example, the stellar fraction is 27\% (compared with 44\% for the
isotropic case $\beta$ = 0). The variation of the density profiles
with varying depths of the gravitational potential ($\psi_0$) and
varying amounts of gas ($\Lambda$) are similar to those shown in
Figure 1 for isotropic models.  

\section{REMOVAL OF THE GASEOUS COMPONENT} 

For the next stage of this calculation, we let all of the gas be
removed from the cluster on a short time scale.  We thus assume that
the distribution function of the stars does not have time to adjust
and hence all of the stars suddenly find themselves in a new
environment with a less deep gravitational potential well, but they
initially retain the same distribution function, as given by equation
[\ref{eq:distfun}] or [\ref{eq:dfani}] evaluated using the {\it old}
values of the potential.  In nature, the removal of gas does not take
place instantaneously, however, so this calculation represents a
limiting case of the physical problem.

The energy required to remove gas from a cluster can be easily obtained. 
The binding energy of a cluster is given by $E = k GM^2/R$, where 
$k$ is a constant of order unity that depends on the cluster shape. 
In terms of representative values, we can write 
\be 
E \approx 4 \times 10^{46} {\rm erg} 
\Bigl( {M \over 1000 M_\odot} \Bigr)^2 
\Bigl( {R \over 1 {\rm pc} } \Bigr)^{-1} \, . 
\ee 
Since a supernova explosion has a typical energy rating of $E_{SN}
\sim 10^{51}$ erg, one supernova provides enough energy to have a
devastating impact on the gaseous component of a cluster. However,
radiative cooling is very efficient in molecular clouds and can
dispose of much of the injected energy on short time scales (Wheeler
et al. 1980; Franco et al. 1994).  In addition, gas dispersal takes
place efficiently even in the absence of supernovae (Palla \& Stahler
2000). As another mechanism, winds from young stellar objects inject a
large amount of mechanical luminosity into the surrounding medium,
with a typical scaling law of $L_{\rm mech}$ $\sim 0.01 L_\ast$ (e.g.,
Lada 1985).  According to this relation, a young stellar object with
$L_\ast$ = 3 $L_\odot$ imparts an energy $\Delta E \sim 10^{45}$ erg
during a time interval $\sim$0.3 Myr. The outflows from YSOs thus
impart enough energy to facilitate gas removal. At the high densities
of molecular clouds, however, gas removal by photoionization and
photodissociation can be a more effective mechanism (Diaz-Miller 
et al. 1998). In any case, the energy to remove gas from clusters 
is easily obtained, but the requisite momentum transfer is more
difficult to achieve. Although the actual gas dispersal mechanism
remains unclear, gas is observed to dissipate in less than $10^7$
years, even for aggregates containing no massive stars (Palla \&
Stahler 2000).

In the models of this paper, we calculate the fraction of stars
remaining in a cluster (after gas removal) as a function of star
formation efficiency. We note, however, that the star formation
efficiency is coupled to the gas removal process. If gas is removed
slowly, then the transformation of gas into stars operates over a
longer period of time and the star formation efficiency can be larger 
(although the opposite is not necessarily true). 

Against this background, we now proceed with the calculation. 
The immediate change $\Delta \psi_{\rm g}$ in the gravitational 
potential -- that just due to gas removal -- is given by 
\be 
\Delta \psi_{\rm g} (\xi) = (\Delta \psi_{\rm g})_{\rt} + \Lambda 
\Bigl[ {1 \over \xit} \atan \xit - {1 \over \xi} \atan \xi + 
{1 \over 2} \ln \bigl( {1 + \xit^2 \over 1 + \xi^2} \bigr) \Bigr] \, , 
\ee 
where the change in potential $(\Delta \psi_{\rm g})_{\rt}$ at 
the outer cluster boundary can also be evaluated to obtain 
\be
(\Delta \psi_{\rm g})_{\rt} = \Lambda \bigl( 1 - 
{1 \over \xit} \atan \xit \bigr) \, . 
\ee
This form for the outer boundary condition assumes that the gaseous 
halo is embedded within a much larger gas cloud with the properties 
of an isothermal sphere. This assumption thus specifies the constant 
$\Phi_0$ = 0. We can combine the above two expressions and simplify 
the result to obtain 
\be 
\Delta \psi_{\rm g} (\xi) = 
\Lambda \Bigl[ 1 - {1 \over \xi} \atan \xi + {1 \over 2} 
\ln \bigl( {1 + \xit^2 \over 1 + \xi^2} \bigr) \Bigr] \, .
\ee
The physical change in the potential is thus given by 
$\Delta \Psi_{\rm g}$ = $\sigma^2 \, \Delta \psi_{\rm g} (\xi)$. 

With the gas removed from the cluster, the stars find themselves in a
new (shallower) potential and hence some stars must leave the system.  
The gravitational potential changes because of both gas removal and 
the removal of the high velocity stars. At a given location within 
the cluster, the original distribution of stars had a range of 
possible velocities given by 
\be 
0 \le v^2 \le 2 \Psi \, . 
\ee
In the new configuration, without the gaseous component, the stars
that remain bound to the cluster must have velocities in the range 
\be
0 \le v^2 \le 2 (\Psi - \Delta \Psi) \equiv \vm^2 \, , 
\ee
where $\Delta \Psi$ includes the change in potential due to stars 
leaving the system, and where we have defined a maximum velocity 
$\vm$ in the final equality.  We thus need to solve for the new 
potential, which we can write as $\phi = \vm^2 / 2 \sigma^2$. 

\subsection{Reduced density profiles for Isotropic models} 

For our model with isotropic velocity dispersion, 
the density of stars remaining is given in terms of $\phi$ 
(or, equivalently, $\vm$) according to the relation 
\be 
\rhostar ({\rm bound}) = \int_0^{\vm} \, 4 \pi v^2 dv 
\, f(\ep) \, = \rho_1 \, \Bigl[ {\rm e}^\psi \erf (\sqrt{\phi}) - 
2 \sqrt{ {\phi/\pi}} \bigl( {\rm e}^{\psi - \phi} + 
{2 \over 3} \phi \bigr) \Bigr] \, . 
\ee
In the above expression, the old potential $\psi$ no longer 
has the physical meaning of the gravitational potential. 
Instead, $\psi (\xi)$ is simply a known function which determines 
the distribution function of stars, only a fraction of which 
will remain in the cluster. The fraction that remains depends 
on the new potential $\phi(\xi)$ which must be determined by solving 
the following new version of the Poisson equation, 
\be 
\xi^{-2} {d \over d\xi} \Bigl( \xi^2 {d \phi \over d\xi} \bigr) 
= - {9 \over g(\psi_0) } \Bigl[ {\rm e}^\psi \erf (\sqrt{\phi}) - 
2 \sqrt{ {\phi/\pi}} \bigl( {\rm e}^{\psi - \phi} + {2 \over 3} 
\phi \bigr) \Bigr] \,  , 
\label{eq:diffeq2} 
\ee
with $\psi(\xi)$ a known function.  To find the new potential, and
hence the new density, we must also specify the boundary condition at
the cluster center $\xi=0$. Here we invoke the condition
\be 
\phi(0) = \psi(0) - \Delta \psi_{\rm g} (0) = 
\psi(0) - {\Lambda \over 2} \ln (1 + \xit^2) \, ,
\ee
which accounts for the change in potential due to gas loss. 

With the problem now completely specified, we can solve for the
original equilibrium structure of the cluster as a function of the
potential well depth $\psi_0$ and the gas parameter $\Lambda$.  We can
then integrate equation [\ref{eq:diffeq2}] to find the density profile
for the stars remaining after the gas is removed.  The mass fractions
are listed in Table 1 for the representative case of $\psi_0$ = 10.
Notice how the results differ from those predicted by equation
[\ref{eq:fsimple}]. For $\Lambda = 1/2$, for example, the total gas
mass is somewhat greater than the original mass in stars. After the
gas is removed, nearly 73\% of the stellar component remains, whereas
the virial argument predicts that none of the stars remain bound.

The difference between these results lies in two key features of this
present treatment: [A] By formulating the problem in terms of the
distribution function $f(\ep)$ for the stars, we take into account the
low velocity tail of the distribution (which always tends to remain
bound). [B] The gaseous halo in our models is tied to the background
molecular cloud, and not directly tied to the stellar component of the
cluster (although the gravitational interaction is included).  This
complication allows the gas density distribution to have a markedly
different form from that of the stars.  In particular, the stellar
component exhibits a relatively steep fall off near the edge of the
cluster, whereas the gas density continues to follow its usual
power-law distribution (see Figure 1).  As a result, the mass of the
gaseous component is concentrated more towards the outer portion of
the cluster (compared to the stellar component) and gas removal has a
less destructive effect on stellar population of the cluster.

Figure 3 shows the fraction $\fun$ of stars remaining after gas
removal as a function of the star formation efficiency, defined as
$\epsilon_\ast$ $\equiv$ $\int \rhostar dV / M$.  Notice that the
shape of these curves is markedly different from the singular behavior 
predicted by equation [\ref{eq:fsimple}].  As a very rough
approximation, the entire collection of curves can be described by the
simple function $\fun$ = $2 \epsilon_\ast - \epsilon_\ast^2$, as shown
by the dashed curve.  Much more accurate fits to individual models can
be obtained. For the $\psi_0 = 10$ model, for example, the dotted
curve shows a fit using a function $\fun = (2 \ewig - \ewig^2)^{2/3}$,
where $\ewig \equiv (10 \epsilon_\ast - 1)/9$ is a stretched variable.

The clusters produced by this rapid gas removal have a truncated
distribution function (by construction). From this state, the cluster
tends to evolve toward a new configuration with a smooth distribution
function. Because this subsequent evolution takes place through
dynamical interactions of the constituent stars, the time scale for
readjustment is longer than the gas removal time (but is nonetheless
shorter than the total dynamical evolution time scale). This longer
term evolution poses an interesting problem for future work.

\subsection{Reduced density profiles for Anisotropic models} 

For the anisotropic models developed in the previous section, 
the new density profile, after the gas is removed from the cluster, 
can be written in the form 
\be
\rhostar = \rho_1 \sqrt{2 \over \pi} \Biggl\{ {\rm e}^{\psi} 
K_\rho (\phi, \alpha) + {\sqrt{2 \phi} \over 2 \alpha^2} - 
{1 \over \alpha} \atan(\alpha \sqrt{2 \phi}) 
\Bigl[ {\rm e}^{\psi-\phi} + \phi + {1 \over 2 \alpha^2} \Bigr] 
\Biggr\} \, . 
\ee
Keep in mind that $\psi$ is the original potential, before gas 
removal, and $\phi$ is the physical potential as determined by 
the solution to the Poisson equation. 

As in the previous case, we can find the equilibrium structure of the
cluster in the presence of a gaseous component, and then use the above
formulation to find the density profile of the stars remaining after
gas is removed. Compared with the isotropic models of the previous
section, the anisotropy of the velocity distribution allows a greater
fraction of the stars to remain after gas removal. For a collection
of models with varying $\psi_0$, Figure 4 shows the fraction $\fun$ of
stars remaining after gas removal as a function of the star formation
efficiency $\epsilon_\ast$ (the solid curves). The dashed curve shows
the function $\fun$ = $(2 \epsilon_\ast - \epsilon_\ast^2)^{1/2}$,
which provides a rough fit to the family of curves as shown. As
illustrated here, the anisotropic models preserve an even greater
fraction of their stars than the isotropic models considered
previously (compare Figure 3 and Figure 4).

\section{COMPARISON WITH OBSERVATIONS} 

The Trapezium cluster is a nearby young cluster in Orion and is often
used an example of a forming bound cluster (e.g., Hillenbrand \&
Hartmann 1998; McCaughrean \& Stauffer 1994; Hillenbrand 1997). Since
this cluster is relatively young (the stars have a quoted mean age of
$\sim 0.8$ Myr), it provides a good point of comparison for the models
of newly formed clusters. The total mass within the central 2.06 pc
region is estimated to be in the range 4500 -- 4800 $M_\odot$
(Hillenbrand \& Hartmann 1998); these same authors estimate that the
core radius $r_0 \approx 0.2$ pc and find a total number of stars
$N_\ast \approx 2300$ within the central $r=2.06$ pc region. Within
this same volume, the total mass of the cluster is larger than that of
the stellar component by a factor of about two, i.e., stars constitute 
40--50\% of the cluster mass within the inner 2 pc.

The characteristics of this cluster are roughly consistent with the
cluster formation scenario outlined in the Appendix. If we use the
isothermal version of the model, for example, we would need an
effective sound speed of $a \approx 2.3$ km/s to form a 4800 $M_\odot$
cluster in 1.7 Myr; these same values imply $r_\infty \approx 2$
pc. In order to obtain a centrifugal radius (and a core radius) of
$R_C = r_0 = 0.2$ pc, the initial rotation rate of the core would have
to be $\Omega_1$ $\approx$ 0.4, which is a reasonable value for
molecular clouds.  Similarly, for the logatropic model, a pressure
scale of $P_0$ $\approx$ 6.6 $\times 10^{-10}$ dyn cm$^{-2}$ would
imply $r_\infty$ = 2 pc and a formation time of 2.9 Myr. To obtain the
same core radius of $r_0 = R_C$ = 0.2 pc, the required rotation rate
is again $\Omega_1 \approx 0.4$.  We thus argue that inside-out
collapse models can be made consistent with this particular observed
young cluster.

Hillenbrand \& Hartmann (1998) have already used lowered isothermal
models to fit to the stellar component of the Trapezium cluster data
and find good agreement using a central potential $\psi_0$ = 9. This
model does not include the gaseous component or the possibility of
anisotropy in the velocity distribution; however, including these
additional parameters can only make the fit better.  We can obtain a
reasonable model fit to this same cluster using $\psi_0$ $\approx$ 10.
For an isotropic model with $\Lambda$ = 1/2, the total mass in mass is
comparable to the mass in stars. If the gas is removed from such a
cluster, nearly 73\% of the stellar mass remains bound, with the
remainder escaping to fill the field with stars.  These models extend
out to a truncation radius $r_T$ which is about 100 core radii and
thus extends many times farther out than the observed portion of the
cluster. In order to have the gas mass and the stellar mass nearly
equal in the (observed) central 2 pc region, we need a larger gas
parameter of $\Lambda \approx$ 1.5; for this case, only about 48\% of
the stars remain bound after the gas leaves the system. After gas
removal, this simple model thus predicts that the central region of
the Trapezium cluster will remain a bound entity with approximately 
$N \approx$ 1000 stars. If the velocity distribution is not isotropic,
but instead becomes more radial in the outer cluster, then an even
larger fraction of stars could be retained after gas removal.

In the existing literature, there are few predictions regarding the
fate of the Trapezium for comparison.  An older study using only 16
stars indicates a half-life of only $10^6$ years (Allen \& Poveda
1975). On the other hand, a recent numerical study (Kroupa 2000) 
obtains results in good agreement with the semi-analytic models of 
this paper. 

\section{SUMMARY and DISCUSSION} 

In this paper, we have used results from star formation theory and
stellar dynamics to develop a working model for the early evolution
of open star clusters. More specifically, we have constructed a
sequence of equilibrium cluster models, including both isotropic and
anisotropic velocity dispersions.  These models represent a
straightforward generalization of previous work to include a gaseous
component (see Figures 1 and 2). We have then considered the effects 
of gas removal from these clusters. 

When a cluster loses its gaseous component, the subsequent evolution
of the stellar component is largely determined by the distribution
function $f(\ep, L)$. The cluster always contains some stars with low
velocities (low kinetic energy), and hence at least some stars always
remain behind.  Because the distribution function depends sensitively
on the velocity, however, the number of stars left behind is sensitive
to the shape of the distribution.  In this class of models, relatively
more stars remain bound to the cluster than suggested by virial
arguments. For example, the model most applicable to young open star
clusters, with anisotropy parameter $\beta = 1$ and with equal mass in
gas and stars, initially retains 93\% of its stars after the gaseous
component is removed.  In general, the fraction $\fun(\epsilon_\ast)$
of stars left behind after gas removal is a smoothly varying function
of star formation efficiency $\epsilon_\ast$ (see Figures 3 and 4).
As a crude approximation spanning the entire space of solutions for
isotropic models, the fraction $\fun$ can be characterized by a
elementary function of the form $\fun$ = $2\epsilon_\ast -
\epsilon_\ast^2$.  More accurate fits can be obtained for specific
cluster models.  Similarly, anisotropic models with $\beta = 1$ can 
be described by the function $\fun$ = 
$(2\epsilon_\ast - \epsilon_\ast^2)^{1/2}$.  

Along the way to the above results, we have developed a simple model
of cluster formation (described in the Appendix). If clusters form out
of the collapse of a large molecular cloud structure, the collapse
flow can be described by scaled-up versions of the infall collapse
solutions for individual star formation events.  The time scale for
individual stars to form is much shorter than the time required for
the cluster to form, and the dynamics on the larger size scale of the
cluster become purely ballistic. The central portion of the forming
cluster, at radii smaller than the centrifugal radius $R_C \sim 0.1$
pc, has time to dynamically relax and tends to exhibit an isotropic
velocity dispersion; the centrifugal radius also sets the core radius
of the cluster. In the outer portion of the forming cluster, $r >
R_C$, the velocity dispersion becomes anisotropic and nearly radial.

This paper represents a modest step toward a unified picture of
cluster formation, early evolution, and dispersal. We have included a
gaseous component in the construction of semi-analytical equilibrium
models of clusters and have considered the initial effects of gas
removal. The subsequent development of the resulting stellar systems
should be studied next.  More detailed models of cluster formation
constitute another fruitful area for further work. With a more
definitive picture of cluster formation in hand, the distribution
function of the stars in the earliest phases can be more precisely
specified and the fraction of stars that remain after gas removal can
be more accurately determined. Another important complication is to
include a different distribution function for stars of different
masses.  Once we understand these issues of cluster formation and
early evolution, we can then determine their effects on the star
formation process.

\acknowledgments

We would like to thank Gus Evrard, Charlie Lada, Greg Laughlin, Phil
Myers, Doug Richstone, and Frank Shu for useful discussions. This work
was supported by funding from The University of Michigan and by
bridging support from NASA Grant No. 5-2869.

\appendix

\section{A Cluster Formation Model}

In this Appendix, we present a simple model for cluster formation. For
robust bound clusters forming within molecular clouds, we expect the
original proto-cluster structure to collapse as a whole. To obtain a
mathematical description of this collapse, we consider the flow that
eventually produces a cluster to be a scaled up version of the
collapse flows that have been studied previously for single star
formation (e.g., Shu 1977; Terebey, Shu, \& Cassen 1984; Jijina \&
Adams 1996). This approach should capture the basic essence of the
collapse problem.  In this case, the collapse of a molecular cloud
region proceeds from inside-out. The central portion of the flow
approaches a ballistic (pressure-free) form and helps determine the
velocity distribution of the forming cluster.  Even for infalling gas,
the inner limit of the collapse flow always approaches pressure-free
conditions; infalling stars are manifestly ballistic.  The time scale
for individual star formation events is $\sim10^5$ years (Myers \&
Fuller 1993; Adams \& Fatuzzo 1996), whereas the time scale for the
entire cluster to form is somewhat longer, 1--3 Myr.  We thus expect
most of the stars to form while the overall collapse of the cluster is
still taking place.  In apparent support of this picture, observations 
suggest that cluster formation takes place within only about one 
sound crossing time of the system (Elmegreen 2000). 

For a given gravitational potential, we find the orbital solutions for
stars (or gas parcels) falling towards the cluster center. In the
standard infall calculation, the inner solution is derived using the
gravitational potential of a point source. Since this potential is
spherically symmetric, angular momentum is conserved and the motion is
confined to a plane described by the coordinates $(r, \phi)$; the
radius $r$ is the same in both the plane and the original spherical
coordinates. The angular coordinate $\phi$ in the plane is related to
the angle in spherical coordinates by the relation $\cos \phi$ = 
$\cos \theta$ / $\cos \theta_0$, where $\theta_0$ is the angle of the 
asymptotically radial streamline (see below).  For zero energy orbits,
the equations of motion imply a cubic orbit solution, 
\be
1 - {\mu \over \mu_0} = \zeta (1 - \mu_0^2) \, . 
\label{eq:orbit} 
\ee
Here, the trajectory that is currently passing through the position 
given by $\zeta$ and $\mu \equiv \cos \theta$ initially made the angle 
$\theta_0$ ($\mu_0$ = $\cos \theta_0$) with respect to the rotation 
axis. The quantity $\zeta$ is defined by 
\be
\zeta \equiv {j_\infty^2 \over G M r} \,
= {R_C \over r} \, ,  
\label{eq:zeta} 
\ee
where $j_\infty$ is the specific angular momentum of parcels of gas
currently arriving at the cluster center along the equatorial plane. 
The second equality defines a centrifugal radius $R_C$.
We assume that the initial cloud is uniformly rotating at a constant
rotation rate $\Omega$, so that $j_\infty = \Omega r_\infty^2$, 
where $r_\infty$ is the starting radius of the material that is 
arriving at the center at a given time. 

To evaluate the radii $R_C$ and $r_\infty$, we invert the mass
distribution of the initial state. For an isothermal cloud, the 
mass profile and the centrifugal radius are given by 
\be
M(r) = {2 a^2 r \over G} \, , \qquad 
r_\infty  = {GM \over 2 a^2 } \, , \qquad {\rm and} \qquad 
R_C = {\Omega^2 G^3 M^3 \over 16 a^8} \, ,
\label{eq:isoform} 
\ee
where $a$ is the isothermal sound speed.  For this case, the infall
collapse solution (Shu 1977) indicates that the flow exhibits 
a well defined mass infall rate $\dot M$ = $m_0 a^3 /G$, where 
$m_0 \approx 0.975$. To form a cluster with mass $M = 1000 M_\odot$ in
1 Myr, for example, the required effective sound speed is $a \approx$
1.63 km/s. With this central mass, the region that originally filled a
volume of radius $r_\infty \approx$ 0.81 pc has fallen to the center
(within $R_C$).  The size of the collapsing region is about twice as
large, $r_H$ = $at$ $\approx$ 1.66 pc, and contains a total mass $M
\approx$ 2000 $M_\odot$. The centrifugal radius $R_C$ $\approx$ 0.1 pc
for $\Omega$ = 1 km s$^{-1}$ pc$^{-1}$ (a typical observed value;
Goodman et al. 1993). This centrifugal radius is comparable to the
expected core radius $r_{\rm core}$ of a newly formed cluster, and we
make the rough identification $r_{\rm core} \sim R_C$.

We also consider initial states for non-isothermal conditions.
Molecular linewidths often show a substantial nonthermal component
with a density dependence of the form $\Delta v$ $\propto \rho^{-1/2}$
(e.g., Larson 1985; Myers \& Fuller 1992).  If we use a ``logatropic''
equation of state $P = P_0 \log\rho/\rho_0$ to describe such a fluid
(Lizano \& Shu 1989; Jijina \& Adams 1996; McLaughlin \& Pudritz 1997; 
Galli et al. 1999), the equilibrium mass distribution and the 
corresponding radii $r_\infty$ and $R_C$ are given by 
\be
M(r) = \Bigl[ {2 \pi P_0 \over G} \Bigr]^{1/2} r^2 \, \qquad 
r_\infty  = M^{1/2} \,  \Bigl[ {2 \pi P_0 \over G} \Bigr]^{-1/4} \, , 
\qquad {\rm and} \qquad R_C = {\Omega^2 M \over 2 \pi P_0} \, , 
\label{eq:logform} 
\ee
where $P_0$ is the pressure scale that determines the amount of
nonthermal support in the cloud.  In this case, the mass infall rate
is time dependent.  The total mass $M(t)$ that falls to the center of
the flow during a time $t$ is given by $M = m_0 t^4 (2 \pi G P_0)^{3/2}$ 
$/16G$, where $m_0 \approx 0.0302$.  During logatropic collapse, most
of the mass in the original cloud region is still on the way down,
rather than at the center. If the cluster encompasses the entire
collapsing region, the total mass is about 30 times that of the
central core. A typical scenario would thus have 100 $M_\odot$ in the
central core and 3000 $M_\odot$ still falling inwards.  For this case,
the required pressure scale $P_0 \approx 8.9 \times 10^{-10}$
dyne/cm$^2$. To obtain a centrifugal radius $R_C \approx$ 0.1 pc 
$\sim$ $r_{\rm core}$ as before, we need $\Omega$ $\approx$ 
3 km s$^{-1}$ pc$^{-1}$.

Both the isothermal and the logatropic models can produce a cluster in
$\sim1$ Myr using reasonable values for the input parameters ($\Omega$
and $a$ or $P_0$). In both cases, the effective transport speed
required to support the initial cloud is comparable to the velocity
dispersion of the resulting cluster. For the range of parameters
discussed above, this velocity scale is 1--2 km/s and is roughly
consistent with the velocity dispersion of observed open clusters.  

For this cluster formation scenario to be consistent with the current
paradigm of star formation (e.g., Shu, Adams, \& Lizano 1987), the
cluster environment cannot greatly disrupt the collapse of smaller
regions that produce individual stars. To fix ideas, we assume that
the large scale collapse of the cluster is determined by logatropic
conditions (eq. [\ref{eq:logform}]) and the collapse of individual
star forming regions is given by isothermal conditions (eq. 
[\ref{eq:isoform}]).  To produce a 1 $M_\odot$ star, for example,
the radial extent of the initial pre-collapse region is $r_\infty$ =
$GM/2a^2$ $\approx$ 0.02 pc. In the central region ($r<1$ pc) that
eventually becomes the cluster, the individual star forming sites do
not greatly interfere with each other as long as the number of stars
does not exceed $N \approx (1/0.02)^3 \sim 10^5$. Similarly, the
collapse of an individual star proceeds largely independent of the
tidal forces.  For a cluster of mass $M_{\rm clust} \approx$ 1000
$M_\odot$ and size $R_{\rm clust} \approx$ 1 pc, the tidal radius
$r_T$ $\sim$ 0.1 pc, which is much larger than the size of an
individual infall region ($\sim$0.02 pc).

We expect that the stars in the newly formed cluster retain some
dynamical memory of the velocity distribution of the collapse flow.
In this flow, streamlines entering the central region do not cross
each other. As long as the infall time is longer than the time scale
for individual star formation events, the core regions that produce
stars will not have a chance to interact. [As an aside, note that 
this cluster formation scenario has an initial transient phase (the 
first $10^5$ yr) in which the infall of the cluster takes place faster 
than individual stars can form. Protostars entering the central region 
during this initial time period thus have an opportunity to interact 
and merge; this activity may contribute to the production of more 
massive stars in the cluster center.]

Given the orbital solution, we can find the velocity fields for 
the collapse flow: 
\be
v_r = - \Biggl( {G M \over r} \Biggr)^{1/2}
\Bigl\{ 2 - \zeta (1 - \mu_0^2) \Bigr\}^{1/2} \, , 
\label{eq:vrad} 
\ee
\be
v_\theta = \Biggl( {G M \over r} \Biggr)^{1/2}
\Biggl\{ {1 - \mu_0^2 \over 1 - \mu^2 } \, (\mu_0^2 - \mu^2) \, 
\zeta \Biggr\}^{1/2} , 
\label{eq:vtheta} 
\ee
\be
v_\varphi = \Biggl( {G M \over r} \Biggr)^{1/2} \, (1 - \mu_0^2) 
\, ( 1 - \mu^2 )^{-1/2} \, \zeta^{1/2} \, . 
\label{eq:vazimuth} 
\ee 
Since $\zeta$, $\mu$, and $\mu_0$ are related through the orbit
equation [\ref{eq:orbit}], the velocity field is completely determined for 
any position $(r, \theta)$.  With the velocity field specified, we can 
find the anisotropy in the flow as a function of radius. We define 
the angular average for both the perpendicular component of the velocity 
field and the radial component 
\be
\langle v_\perp^2 \rangle = \langle v_\theta^2 + v_\phi^2 \rangle = 
{G M \over r} \zeta \, I_v \,  \qquad {\rm and} \qquad 
\langle v_r^2 \rangle = {GM \over r} \bigl[ 2 - \zeta I_v \bigr] \, , 
\label{eq:vperp} 
\ee
where the integral $I_v$ is given by 
\be
I_v \equiv \int_0^1 \, d\mu \, (1 - \mu_0^2) \, . 
\label{eq:idef} 
\ee

To evaluate the integral $I_v$, we change the integration variable
from $\mu$ to $\mu_0$ and change the lower end of the range of
integration from 0 to a critical value $\muck$. This difference arises
because streamlines from all initial angles cannot fall to arbitrarily
small radii.  For large radii, streamlines from all values of $\mu_0$
are represented.  Inside the centrifugal barrier $R_C$, however, only
streamlines originating preferentially from the poles reach these
smaller radii.  The last streamline that reaches a given radius is
determined by $\muck$. Evaluating the integral $I_v$, we find 
\be
I_v = \bigl( {2 \over 3} - {4 \over 15} \zeta \bigr) (1 - \muck) + 
\muck \bigl( {2 \over 15} + {4 \over 15} {1 \over \zeta} \bigr) \, . 
\label{eq:i2} 
\ee
This expression simplifies in the inner and outer regimes. For large 
radii, $r \gg R_C$, $\muck \to 0$, and $I_v \to (2/3) (1 - 2 \zeta/5)$. 
In the opposite limit of small radii, $r \ll R_C$, $\muck \to$ 
$1 - 1/2\zeta$, and $I_v \to 8/15 \zeta$. 

In the context of cluster formation, we evaluate the anisotropy of the
flow in the outer regime and assume that the cluster retains some
memory of its initial velocity distribution.  Combining equations 
[\ref{eq:vperp} -- \ref{eq:i2}], we find 
\be
{\cal R}_v \equiv 
{\langle v_\perp^2 \rangle \over \langle v_r^2 \rangle} = 
{\zeta I_v \over 2 - \zeta I_v} = 
{\zeta \over 3} {(1 - 2 \zeta/5) \over (1 - \zeta/3)} \, . 
\ee
To leading order, we thus obtain ${\cal R}_v \sim \zeta = R_C/r$. For
large radii the velocities become nearly radial, whereas for small
radii the velocities become more isotropic. Scattering of the newly
formed stars will be relatively efficient at small radii and can drive
the velocity dispersion even further towards isotropy.  More
specifically, the relaxation time can be written in the form $t_{\rm
relax} \approx 14$ Myr ($r$/1pc)$^2$ ($r_\infty$/1pc)$^{-1}$, where we
have assumed a typical robust cluster with $N$ = 1000 stars. For small
radii $r < 0.1$ pc $\approx R_C$, the relaxation time is less than
about 0.2 Myr.  As a result, the region within the centrifugal radius
will experience several relaxation times during the expected time
scale for the cluster to form.

In this picture, clusters form with a nearly isotropic velocity
dispersion on the inside and a highly radial velocity dispersion on
the outside, with the centrifugal radius $R_C$ providing the boundary
between the two regimes.  Furthermore, the core radius of the cluster
is determined by $r_0 \approx R_C$ $\approx$ 0.1 pc (for typical
initial conditions).

\clearpage 
\begin{deluxetable}{rrrrrrrr} 
\tablecolumns{6} 
\tablewidth{0pc} 
\tablecaption{Parameters for Isotropic models with $\psi_0$ = 10} 
\startdata 
$\Lambda$ & $\xit$ & $m_\ast$ & $m_{\rm gas}$ & $m_{\ast new}$ 
& ${\cal F}_\ast$ \\
\hline 
 0.00 & 224  &   126  &   0.00 &   126  &  1.000 \\
 0.25 & 150  &   84.9 &   51.7 &   73.4 &  0.865 \\ 
 0.50 & 106  &   58.9 &   73.0 &   42.8 &  0.727 \\ 
 0.75 & 78.1 &   42.7 &   80.1 &   26.9 &  0.631 \\ 
 1.00 & 59.6 &   32.5 &   81.0 &   18.5 &  0.569 \\ 
 1.25 & 47.1 &   25.8 &   79.4 &   13.6 &  0.526 \\ 
 1.50 & 38.3 &   21.2 &   77.0 &   10.4 &  0.489 \\ 
 1.75 & 32.0 &   17.9 &   74.3 &   8.09 &  0.451 \\ 
 2.00 & 27.2 &   15.5 &   71.8 &   6.36 &  0.412 \\ 
 2.25 & 23.6 &   13.6 &   69.4 &   4.99 &  0.369 \\ 
 2.50 & 20.8 &   12.0 &   67.2 &   3.88 &  0.323 \\ 
 2.75 & 18.5 &   10.8 &   65.3 &   2.96 &  0.274 \\ 
 3.00 & 16.7 &   9.76 &   63.5 &   2.19 &  0.225 \\ 
 3.25 & 15.1 &   8.90 &   61.9 &   1.55 &  0.174 \\ 
 3.50 & 13.9 &   8.16 &   60.4 &   1.00 &  0.123 \\ 
 3.75 & 12.8 &   7.53 &   59.1 &   0.52 &  0.069 \\ 
 4.00 & 11.9 &   6.97 &   57.9 &   0.09 &  0.013 \\ 
\enddata 
\end{deluxetable}

\clearpage
 
\centerline{\bf FIGURE CAPTIONS} 

\figcaption[]{
Equilibrium density profiles for clusters with isotropic
velocity distributions. The density is scaled to the central density
$\rho_0$ of the stellar component and the radius is given in units 
of the corresponding King radius, i.e., $\xi$ = $r/r_0$.  
(a) A typical density profile with $\psi_0$ = 10 and $\Lambda$ 
= 1/2 (44\% stars and 56\% gas). The dashed curve shows the stellar 
component, the dotted curve shows the gaseous component, and the 
solid curve shows the total density. The dot-dashed curve shows 
the stellar component remaining after gas is removed from the system. 
(b) Collection of density profiles with varying depths of the 
gravitational potential and fixed gaseous halo with $\Lambda$ = 1/2. 
The various curves use $\psi_0$ = 6, 8, 10, 12, and 14. As the depth 
$\psi_0$ increases, the outer radius of the stellar component of the 
cluster grows accordingly (see also Table 1). 
(c) Collection of density profiles for $\psi_0$ = 10 and varying 
amounts of gas, determined by $\Lambda$ = 0, 0.5, 1.0, 1.5, and 2.0.
As $\Lambda$ increases, the outer radius of the cluster decreases. 
\label{fig1}}

\figcaption[]{
Equilibrium density profile for a cluster with an anisotropic velocity
distribution. The density and radius have the same units as in Figure
1.  This model uses $\psi_0$ = 10, $\Lambda$ = 1/2, and $\beta = 1$.
The dashed curve shows the stellar component, the dotted curve shows
the gaseous component, and the solid curve shows the total. The
dot-dashed curve shows the stellar component remaining after gas is
removed from the cluster.
\label{fig2}}

\figcaption[]{
Fraction $\fun$ of stars remaining bound to a cluster after
gas removal for systems with isotropic velocity distributions. The 
solid curves show the fraction $\fun$ as a function of star formation
efficiency $\epsilon_\ast$ for varying $\psi_0$ (here, $\psi_0$ = 6, 8,
10, 12, and 14). The dashed curve shows the function $\fun$ = 2
$\epsilon_\ast - \epsilon_\ast^2$, which provides a rough approximation
to the entire family of curves. The dotted curve shows a more accurate
fit (to the $\psi_0$ = 10 model) using the function $\fun = (2 \ewig -
\ewig^2)^{2/3}$, where $\ewig \equiv (10 \epsilon_\ast - 1)/9$ is a
stretched variable.  
\label{fig3}}

\figcaption[]{
Fraction $\fun$ of stars remaining bound to a cluster after gas
removal for systems with anisotropic velocity distributions. Curves
show the fraction $\fun$ as a function of star formation efficiency
$\epsilon_\ast$ for varying $\psi_0$ (here, $\psi_0$ = 8, 10, and
12). This family of curves was calculated using velocity anisotropy
$\beta$ = 1.  The dashed curve shows the function $\fun$ = 
$(2 \epsilon_\ast - \epsilon_\ast^2)^{1/2}$, which provides a rough 
fit to the family of curves as shown. 
\label{fig4}}

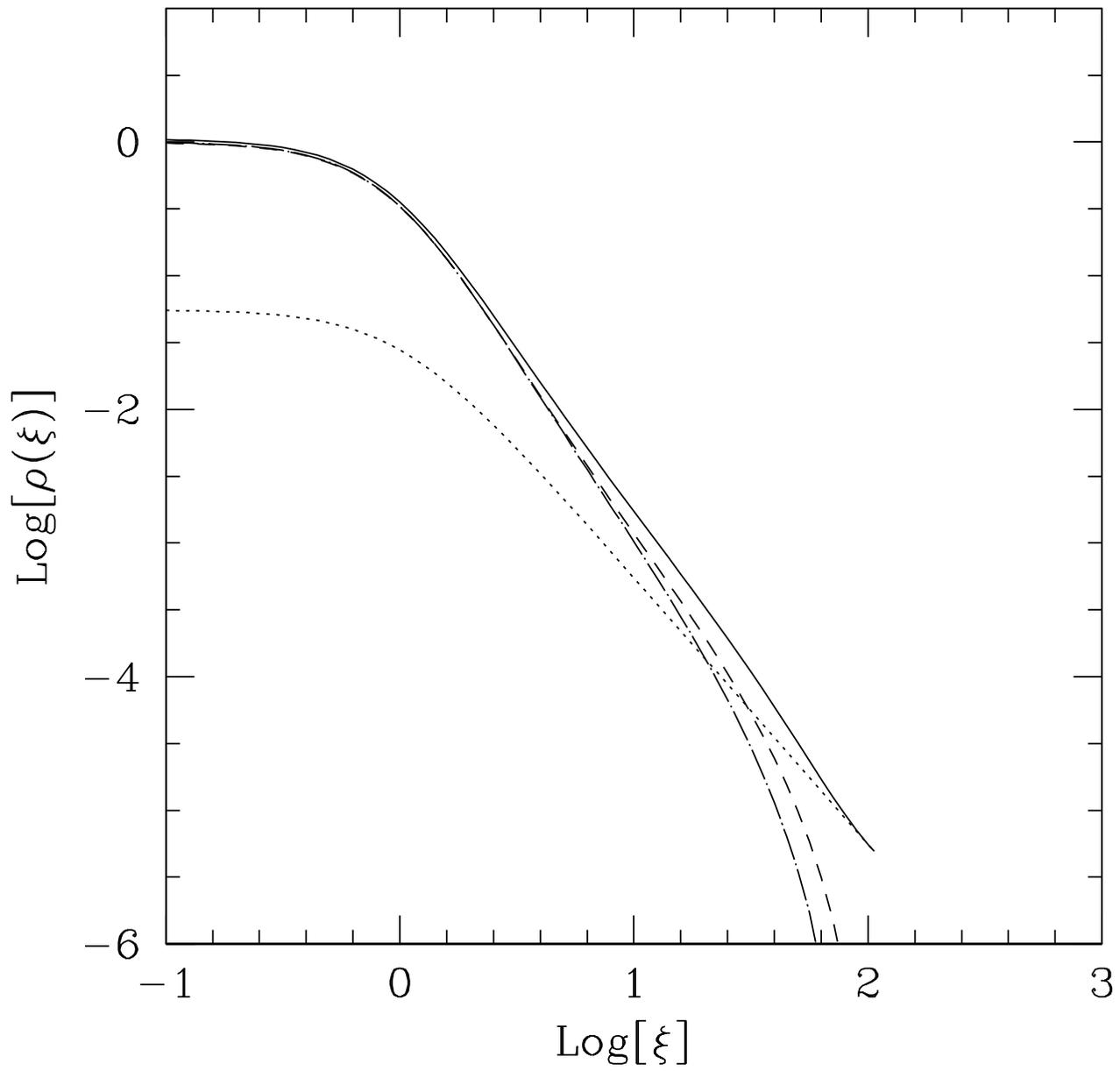
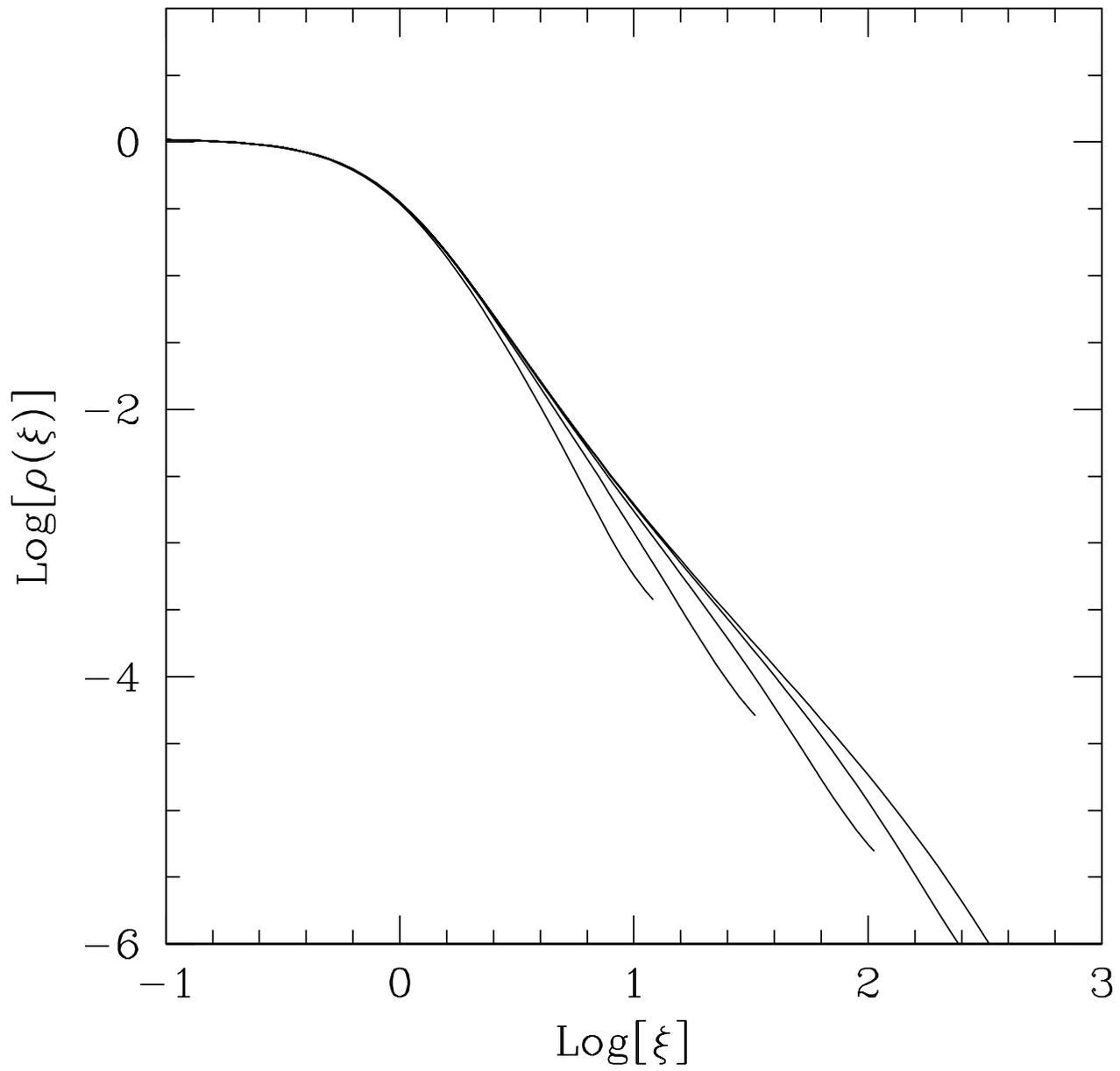
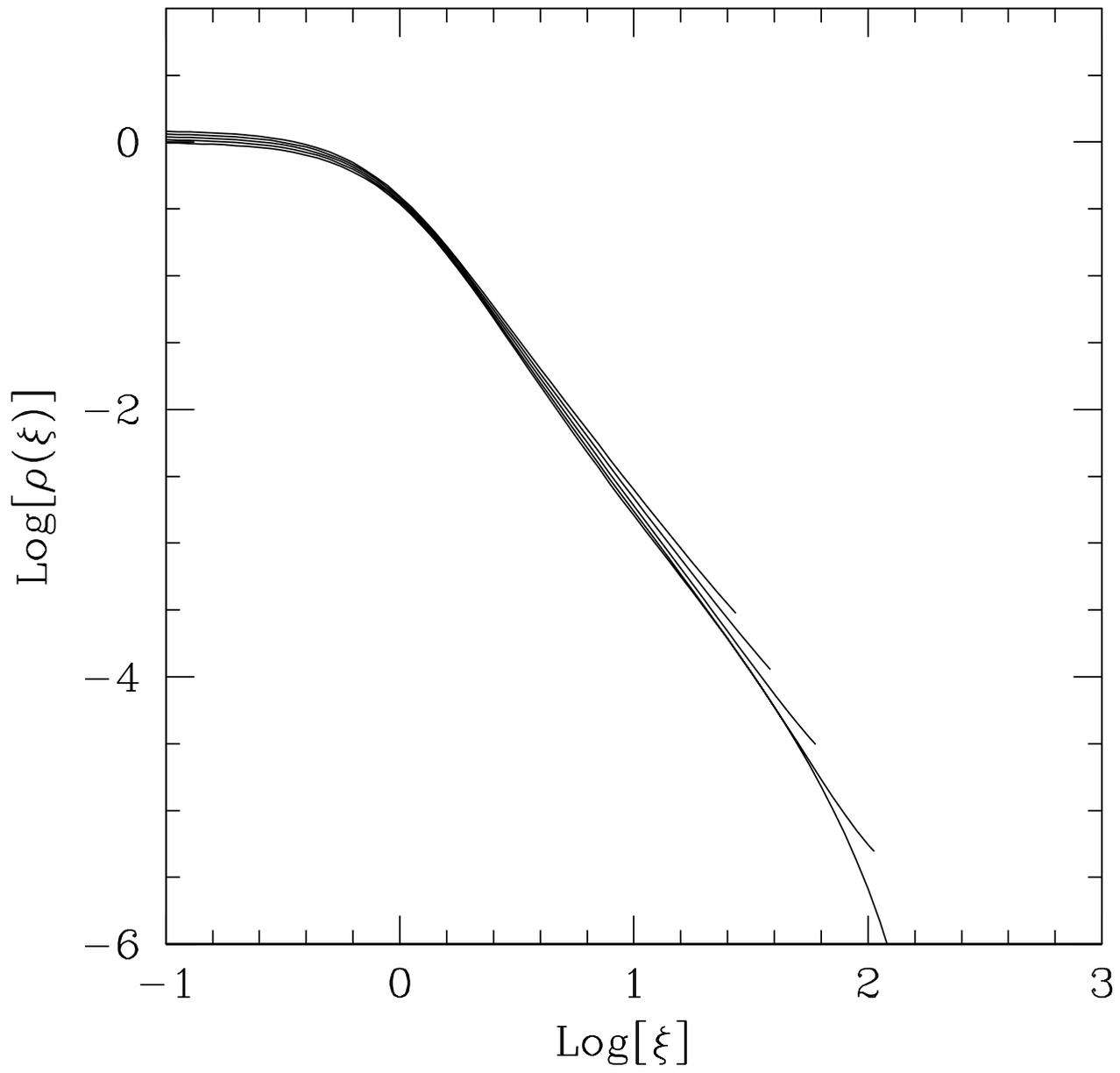
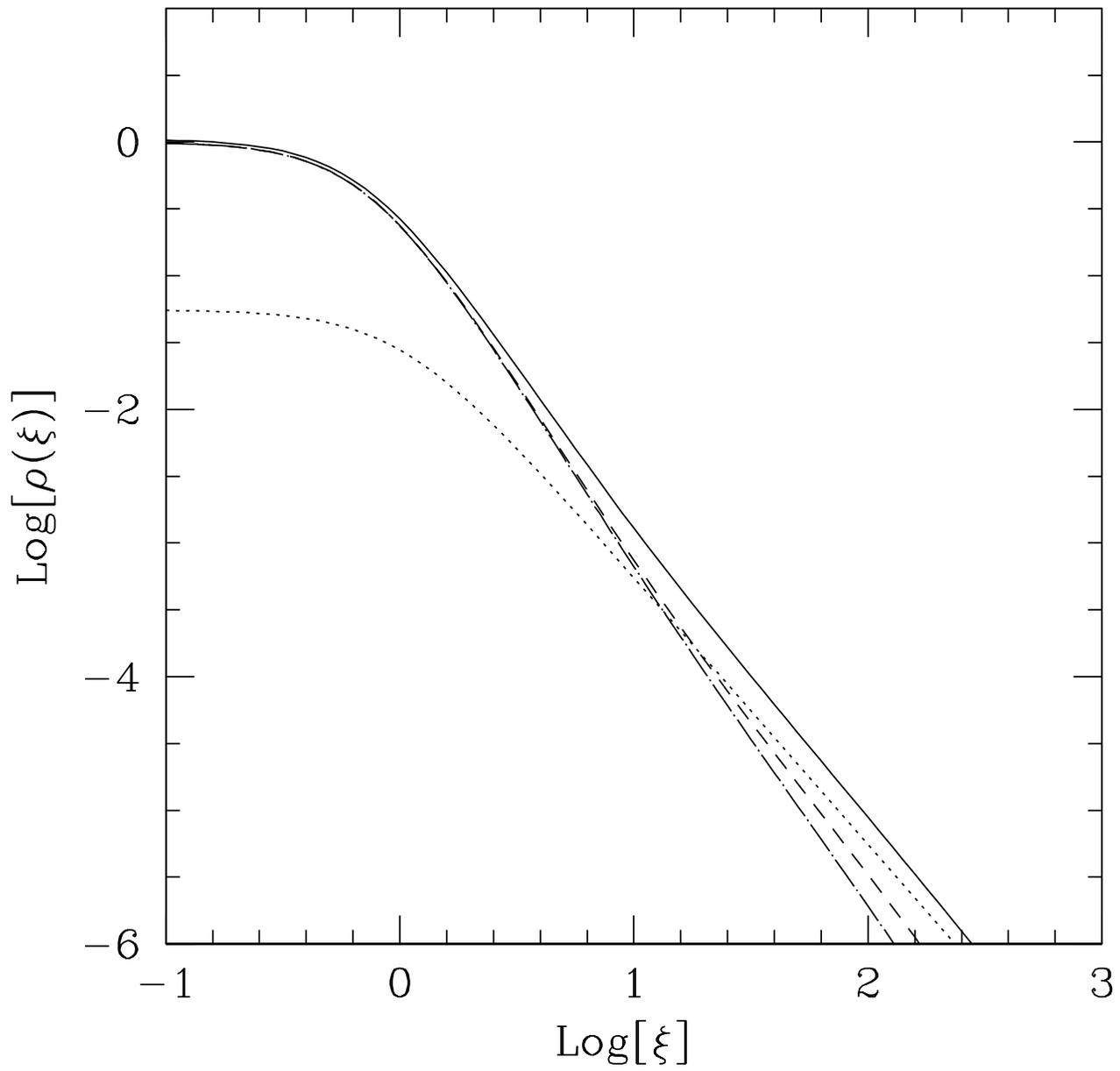
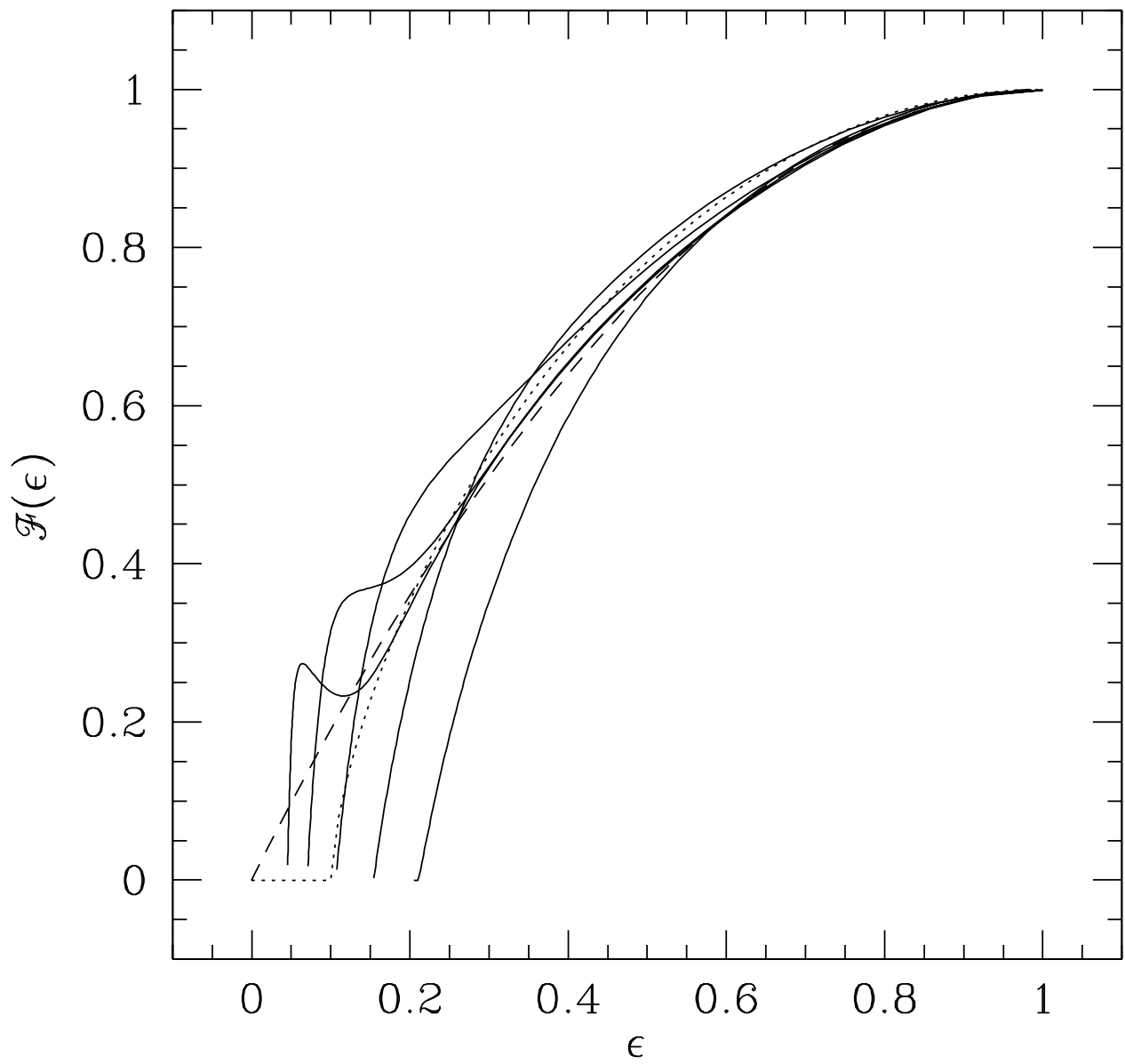
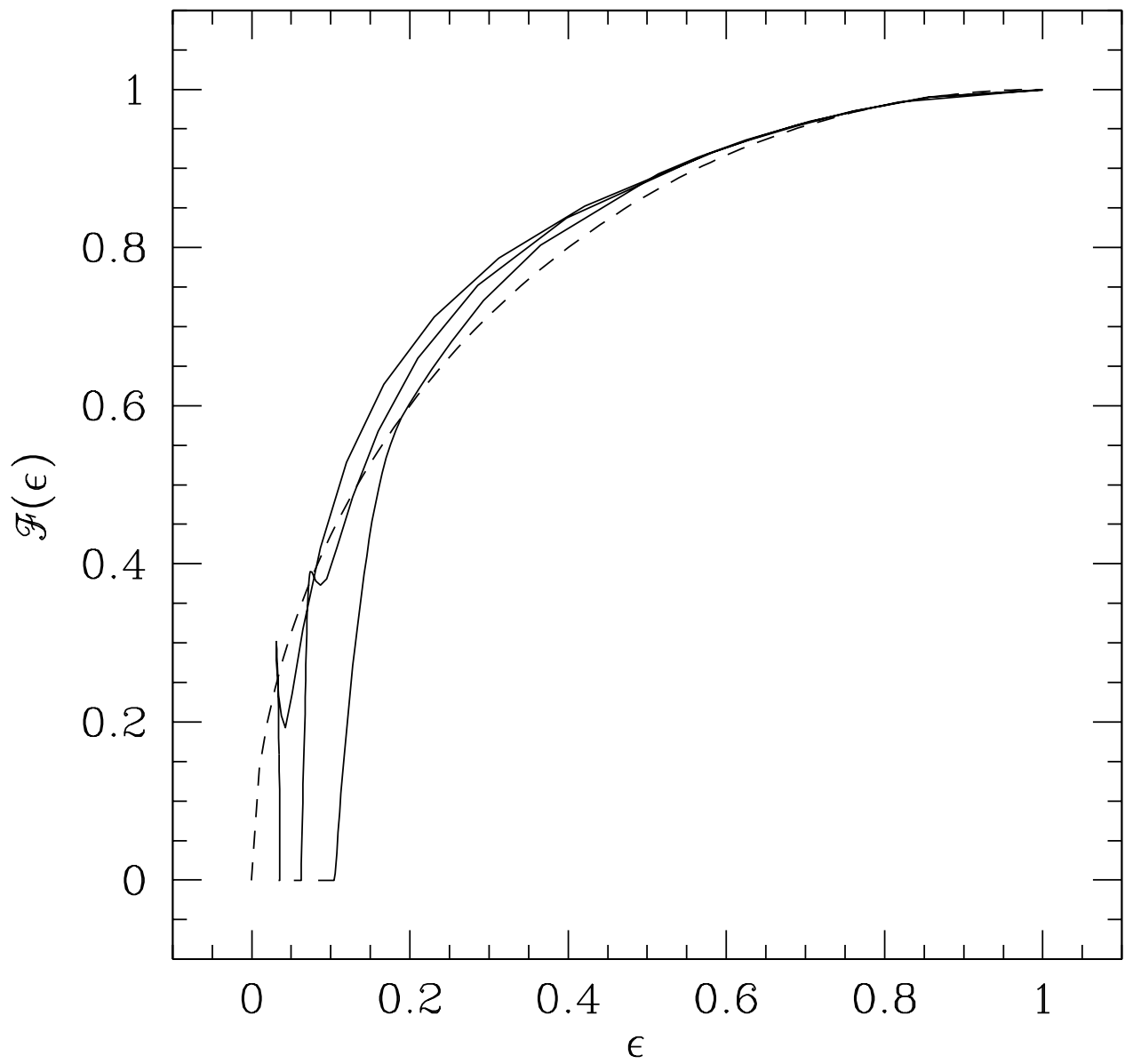
\end{document}